# Band Tail States: Speed Bump or End of the Road for Two-Dimensional Semiconductors?

Zherui Han[1], Robert K. A. Bennett[1], Davide Donadio[2,†], Maritha A. Wang[3], Tara Peña[1], Mikaël Cassé[4], Martin Vanbrabant[5], Valeriya Kilchytska[5], Jean-Pierre Raskin[5], and Eric Pop[1,3,6,*]

[1] *Department of Electrical Engineering, Stanford University, Stanford, CA 94305, USA*
[2] *Department of Chemistry, University of California, Davis, Davis, CA 95616, USA*
[3] *Department of Materials Science and Engineering, Stanford University, Stanford, CA 94305, USA*
[4] *Univ. Grenoble Alpes, CEA, Leti, F-38000 Grenoble, France*
[5] *Institute of Information and Communication Technologies, Electronics and Applied Mathematics (ICTEAM), Université Catholique de Louvain, 1348 Louvain-la-Neuve, Belgium*
[6] *Department of Applied Physics, Stanford University, Stanford, CA 94305, USA*

Contact: [*]epop@stanford.edu, [†]ddonadio@ucdavis.edu

**Abstract:** We propose that the path to realizing the full potential of two-dimensional (2D) electronics lies in understanding and mastering their energy band tail states (BTS). Unlike silicon, which benefits from sharp band edges (Urbach energy $E_U$ ~ 10 meV), today's monolayer 2D semiconductors have rough band edges ($E_U$ ~ 100 meV) caused by spatial potential inhomogeneity. Comparing a physics-based electrical model to experiments, we demonstrate that BTS cause reduced effective mobility and "slow" transistor turn-on with excessive gate voltage in monolayer 2D transistors, which sets a fundamental limit for their energy-efficient operation. We also show that BTS effects in 2D transistors could be mitigated with higher gate capacitance, with strain engineering, and by using bilayer instead of monolayer 2D channels, as long as $E_U$ is reduced below ~ 50 meV. This work provides a theoretical framework and design guidelines to navigate these fundamental limits, and to enable successful integration of 2D semiconductors into low-power nanoelectronics.

## Introduction

Ultrathin, monolayer two-dimensional (2D) semiconductors represent the ultimate thickness limit for a variety of (opto-)electronics applications. For example, in transistors [1] such 2D channels maintain reasonable charge carrier mobility at sub-nanometer thickness and enable better electrostatic control by their gates [2]. In contrast, silicon films with comparable thickness suffer from drastic performance degradation due to surface scattering of charge carriers [2–4]. An important question concerns the fundamental limits of charge transport in 2D semiconductors and their devices, especially when supported or surrounded by insulators. While previous efforts have been devoted to understanding intrinsic charge transport in monolayer 2D semiconductors [5–8], the inherent "thinness" of these materials can cause strain inhomogeneities [9], and their interaction with substrates and dielectrics can also introduce spatially-varying potential landscapes [9–11].

In this work, we analyze the fundamental limitations imposed on 2D electrical transport and transistor operation by such ultrathin material inhomogeneities. We show that thermal fluctuations and spatial inhomogeneities lead to blurring of the conduction and valence band edges into band tail states (BTS). If large enough, these BTS can localize charge carriers, degrade their mobility, and diminish the on-state current of a transistor. However, even moderate BTS reduce the band edge sharpness and limit how steeply a transistor turns on and off, increasing its operating voltage and therefore the power consumed. Here we show that lack of band edge sharpness sets fundamental limits on the electrical behavior of ultrathin 2D semiconductors, even for materials that are otherwise defect-free. We also propose a few solutions which could prevent BTS from being the end of the road for 2D transistors.

First analyzed by Urbach [12] in three-dimensional semiconductors, BTS (or Urbach tail) arise from static and dynamic disorder that distorts the density of states (DOS) into a "tail" below (above) the conduction (valence) band [13,14]. Even ultra-clean, crystalline materials with no impurities and defects have BTS due to intrinsic electron-phonon (e-ph) coupling [15] which broadens the electron self-energy, and due to random potential fluctuations [16] which perturb the local electronic structure.

Realistic 2D semiconductor transistors must use insulators in the gate stack, either as crystalline materials (*h*-BN, $CaF_2$) or, more commonly, amorphous oxides (a-$SiO_2$, a-$HfO_2$). These can introduce spatially inhomogeneous strain and charge variations in the atomically thin 2D semiconductor, causing local potential fluctuations [see **Fig. 1(a,b)**] and BTS [10,17,18]. However, the BTS physics, their magnitude, and their effect on charge transport have been often overlooked in the 2D semiconductor literature. As we discuss below, substantial BTS exist especially in monolayer 2D semiconductors, with the potential to severely impact their transistor operation. Thus, here we address a serious but previously under-explored question on the use of 2D semiconductors in future electronics: what are the limits of charge transport in 2D semiconductors in light of BTS?

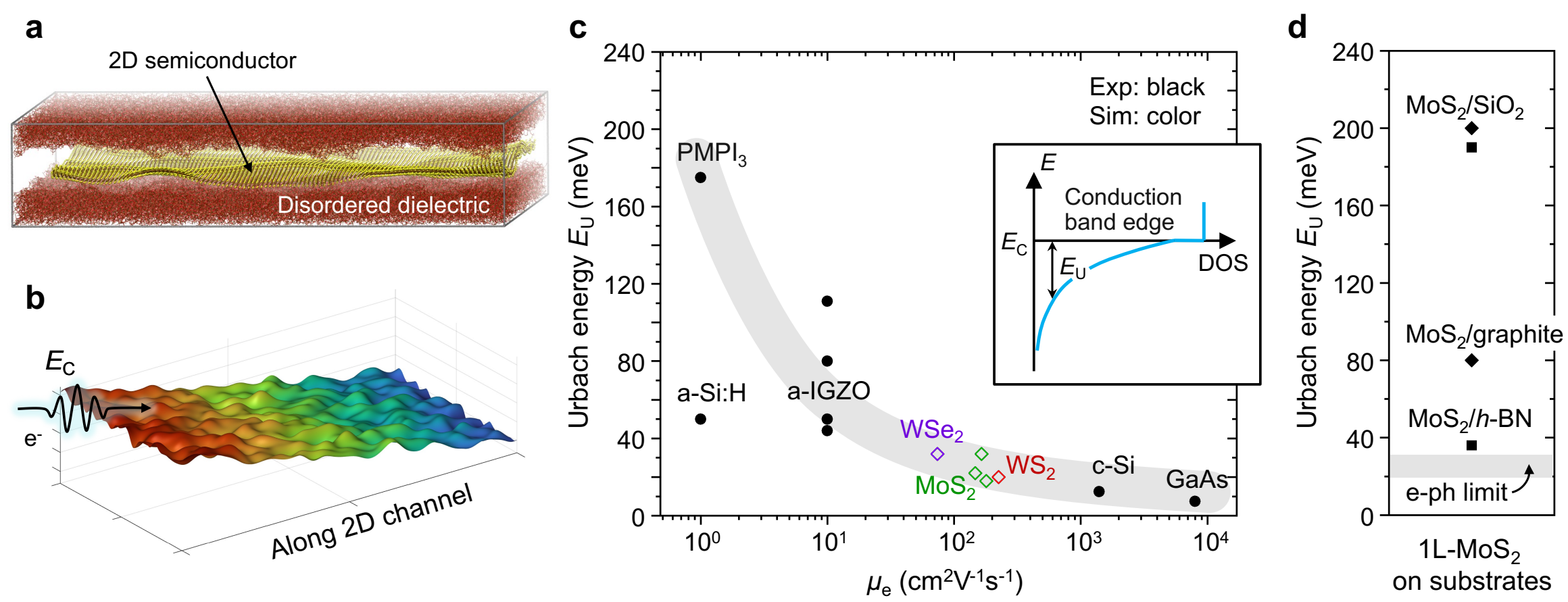

**FIG. 1. Band tail states (BTS) in semiconductors. (a)** Physical origin of BTS in a monolayer 2D semiconductor encased by disordered insulators (here $MoS_2$ encased by a-$SiO_2$). The image is extracted from one snapshot of molecular dynamics simulation (the slab shown here is 30 nm long and 6.5 nm thick), at room temperature. **(b)** Schematic of conduction band ($E_C$) variation along a 2D channel, at low lateral field. An electron wave packet is also illustrated. **(c)** Conduction band Urbach energy $E_U$ in some common semiconductors at room temperature vs. electron mobility $\mu_e$. Filled black symbols are from experiments [14,19–25], unfilled colored symbols are from simulations of intrinsic monolayer TMDs [26–28]. Shaded area is a trend line to guide the eye. Crystalline silicon is marked by c-Si. Electron mobility of amorphous materials (a- prefix) only marks the order of magnitude due to varying quality of materials. Inset shows schematic of exponential BTS density of states (DOS) below the conduction band. **(d)** Experimentally measured $E_U$ for monolayer $MoS_2$ on different substrates; diamond symbols from Ref. [17], square symbols from Ref. [29]. The e-ph limit is the intrinsic range, without a substrate.

## Origins and Magnitude of BTS

The exponential BTS density of states below the conduction band is captured by $g_U(E) = N_U/E_U \exp[-(E - E_C)/E_U]$, where $E_C$ is the conduction band edge, $N_U$ is the total BTS density, and $E_U$ is the Urbach energy, i.e., the exponential 'tail' of the BTS distribution [see the inset of **Fig. 1(c)**]. Although we focus on *n*-type monolayer molybdenum disulfide (1L-$MoS_2$) here as a prototypical 2D semiconductor [2], our analysis can be extended to the valence band and/or to other 2D semiconductors.

In **Fig. 1(c)** we summarize the measured $E_U$ of common bulk semiconductors at room temperature, including GaAs [19], silicon [20–22], amorphous InGaZnO (IGZO) [23,24], and $PMPI_3$, an organic semiconductor [25]. There is an inverse trend between $E_U$ and mobility, with $E_U$ also being higher for disordered materials; this is expected because BTS are an expression of energetic disorder (see, e.g., a-Si vs. c-Si). Even among crystalline materials, $E_U$ scales with the e-ph strength, explaining why GaAs has the highest mobility and lowest $E_U$. This is understandable because e-ph coupling limits electron mobility and simultaneously broadens the electron self-energy at the band edge.

The intrinsic $E_U$ of 2D transition metal dichalcogenides (TMDs) has not been deduced from optical measurements because strong excitonic effects obscure the observation of BTS. Thus, in **Fig. 1(c)** we display their theoretical lower limits at room temperature, as given by e-ph coupling; the three values for intrinsic 1L-$MoS_2$ arise from Green's functions ($E_U \approx 32$ meV) [26], first-principles calculations of electron linewidth (18 meV) [27], and Monte Carlo transport simulations (22 meV) [28] at or close to intrinsic carrier density (see Supporting Information Section I for details). This intrinsic limit of $E_U$ for 1L-$MoS_2$ is larger than for c-Si (~ 10 meV), which is consistent with the lower mobility and stronger e-ph coupling (due to low-energy phonons and high electron DOS near the band edge [30]) in such 2D semiconductors vs. intrinsic bulk silicon.

While the intrinsic BTS limit is set by e-ph coupling, in practical applications 2D materials are placed on or encased by insulating substrates. Recent scanning tunneling spectroscopy (STS) measurements have found substantial BTS in 1L-$MoS_2$ transferred to a-$SiO_2$, with $E_U$ up to 200 meV [17,29], even higher than disordered bulk semiconductors [**Fig. 1(d)**]. In contrast, for $MoS_2$ placed on flat surfaces, $E_U$ is reduced below 80 meV on graphite [17] and approaches the e-ph limit on $h$-BN (estimated ~36 meV from Ref. [29], see Supporting Information Section I for details). We note there is some uncertainty in these $E_U$ estimates, in part due to how the samples were processed (all are layer-transferred $MoS_2$, not directly grown), and in part due to how the $E_U$ was estimated; however, the trends are evident within each study [17,29].

Based on these observations, we suggest that the substrate or environment causes spatial inhomogeneity and disorder in the ultrathin 2D semiconductor. Part of this may be due to interface disorder introduced during nanofabrication [31], and part of it due to dielectric impurities [32,33], dangling bonds [18], and surface roughness which induces nanoscale strain inhomogeneity. Among these, the first two may be eliminated by nanofabrication improvements, but the others are more fundamental and intrinsic to the materials being used. From a mechanical point of view, a 2D monolayer is susceptible to substrate roughness due to its lack of out-of-plane stiffness. This creates a fluctuating potential, which introduces both BTS [13,16] and a spatially-varying band gap through the surrounding dielectric disorder, as shown by exciton spectroscopy experiments [11].

**BTS Effect on Transistors**

We now turn to evaluate the implications of BTS in 2D semiconductor transistors. To this end, we modify an electrical transport model [34] to include the exponential tail DOS, applying it both to 2D and conventional silicon transistors. The drain current normalized by channel width is:

$$I_D = \frac{1}{L}\left[\mu_{band} \int q n_{band}\, dV_{DS} + \int dV_{DS} \int \mu_{hop}(E) g_U(E) f(E) dE\right]$$

where $L$ is the channel length, $q$ is the elementary charge, $n_{band}$ is the mobile charge density (assuming constant DOS above the 2D channel band edge), $V_{DS}$ is the drain-to-source voltage, and $f(E)$ the Fermi-Dirac distribution. This model assumes band-like transport and hopping transport when the Fermi level $E_F$ is above and inside the BTS, respectively, with $\mu_{band}$ and $\mu_{hop}$ the electron band mobility and hopping mobility, respectively. The model also assumes negligible contact resistance, meaning it is best compared with experimental data on few-micron channel lengths among today's 2D transistors [31,35]. Because 2D semiconductors are largely crystalline (unlike, e.g., amorphous semiconductor oxides [36]), this treatment is similar to the multiple-trap-and-release (MTR) model [37], and our calculations of charge transport in the BTS [38,39] show the hopping mobility is quite low, $\mu_{hop} < 0.1$ $cm^2V^{-1}s^{-1}$, with minimal contributions to current (see Supporting Information Section II).

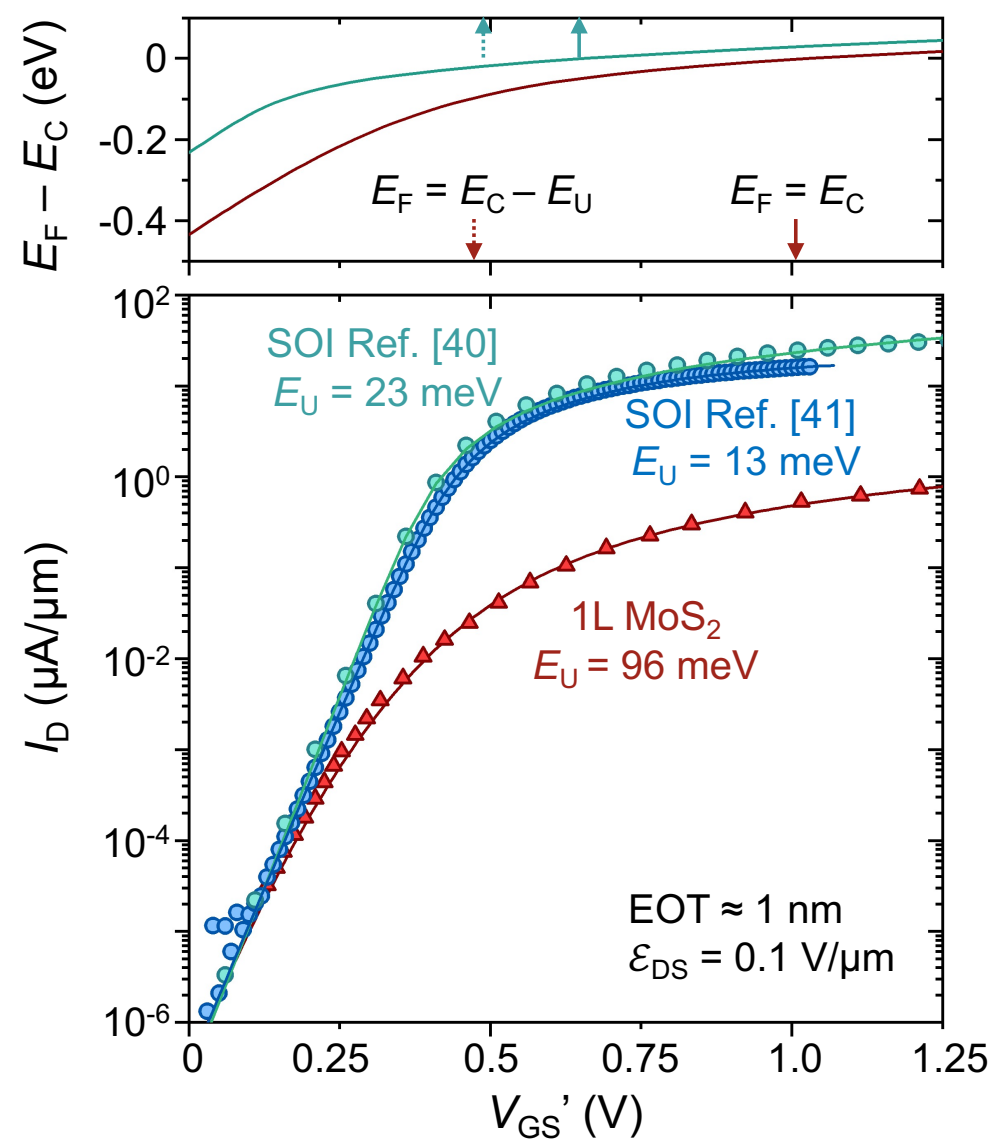

**FIG. 2. Estimated $E_U$ from SOI and $MoS_2$ transistors.** Symbols mark drain current vs. gate voltage measurements of long-channel monolayer $MoS_2$ (red, [31]) and ultrathin SOI (cyan and blue, with silicon channel thickness of 11 and 7 nm, based on device fabrication described in Refs. [40,41], respectively). All devices have metal top-gates with EOT ≈ 1 nm and their current is normalized to the same lateral field $\mathcal{E}_{DS} = V_{DS}/L = 0.1$ V/µm. Lines are our model fitting, yielding $E_U \approx 13$ and 23 meV for SOI and ~ 96 meV for monolayer $MoS_2$. Due to slightly different threshold voltage, the horizontal axis is offset ($V_{GS}$') such that all transistors start at 10 pA/µm for the same voltage. Upper panel shows the simulated position of Fermi level for SOI (cyan) and $MoS_2$ (red) devices, as a function of gate voltage. Small arrows mark where the Fermi level enters the BTS region ($E_F = E_C - E_U$) and conduction band edge ($E_F = E_C$).

In **Fig. 2**, we compare our model (lines) with electrical measurements (symbols) of long-channel monolayer $MoS_2$ [31] and silicon-on-insulator (SOI) transistors [40,41], using just $\mu_{band}$, $E_U$ and $N_U$ as fitting parameters. In principle, one can consider mid-gap defect states [42], but we find their effects do not influence the $I_D$ range in our fitting. All devices in **Fig. 2** have $SiO_2$ substrates, and all have metal top-gates on an equivalent oxide thickness EOT ≈ 1 nm, which is important for direct comparison. They are all also long-channel devices, measured in the linear region at low $V_{DS}$; all devices are normalized to 0.1 V/µm lateral field. For example, the $MoS_2$ device (SOI from Ref. [41]) was at 3× lower (10/3 higher) field, so we increased its current by 3× (decrease its current by 10/3) in the figure. This allows the most direct comparison of a recent monolayer $MoS_2$ device vs. mature silicon technology, without the complications of high-field transport, self-heating, or contact resistance effects.

By fitting our model to these data, we estimate $E_U \approx 96$ meV for the monolayer $MoS_2$ [31] and ~ 10 to 20 meV for the SOI devices at room temperature. Compared to SOI, this top-gated $MoS_2$ transistor is not industrially optimized, and the top-gate deposition process could introduce some defects in the channel, but this comparison is one of the most direct available, with similar top-gate EOT, and amorphous $SiO_x$ above and below the channel. In other words, this is expected to provide a good estimate of the BTS induced in $MoS_2$ by such an amorphous dielectric. We also fit our model to a *back*-gated monolayer $MoS_2$ device on 90 nm $SiO_2$ [35], estimating $E_U \approx 163$ meV (Supporting Information **Fig. S2**).

We do not expect perfect fitting and parameter extraction of the BTS here, because subtle effects of trap states in the oxide dielectrics are not captured. Nevertheless, these are likely minimized in the devices with small EOT used here, and the $E_U$ parameters estimated are in good agreement with STS observations for $MoS_2$ on $SiO_2$ [17,29] and with the Urbach tail expected from optical measurements of crystalline silicon [20,21].

From **Fig. 2**, it is clear that long-channel silicon transistors have a 'sharper' turn-on than long-channel monolayer 2D transistors with the same EOT, and our model fitting shows that the turn-on region is related to their relative magnitude of $E_U$, or how large their BTS region is. In other words, a large BTS region means the transistor requires extra gate voltage to achieve the same level of current compared to the device with sharper band edge. Just like contact resistance effects limit the useful drain-to-source voltage ($V_{DS}$) range, the non-negligible BTS set a fundamental limit on the minimum gate voltage ($V_{GS}$) range of a monolayer $MoS_2$ transistor.

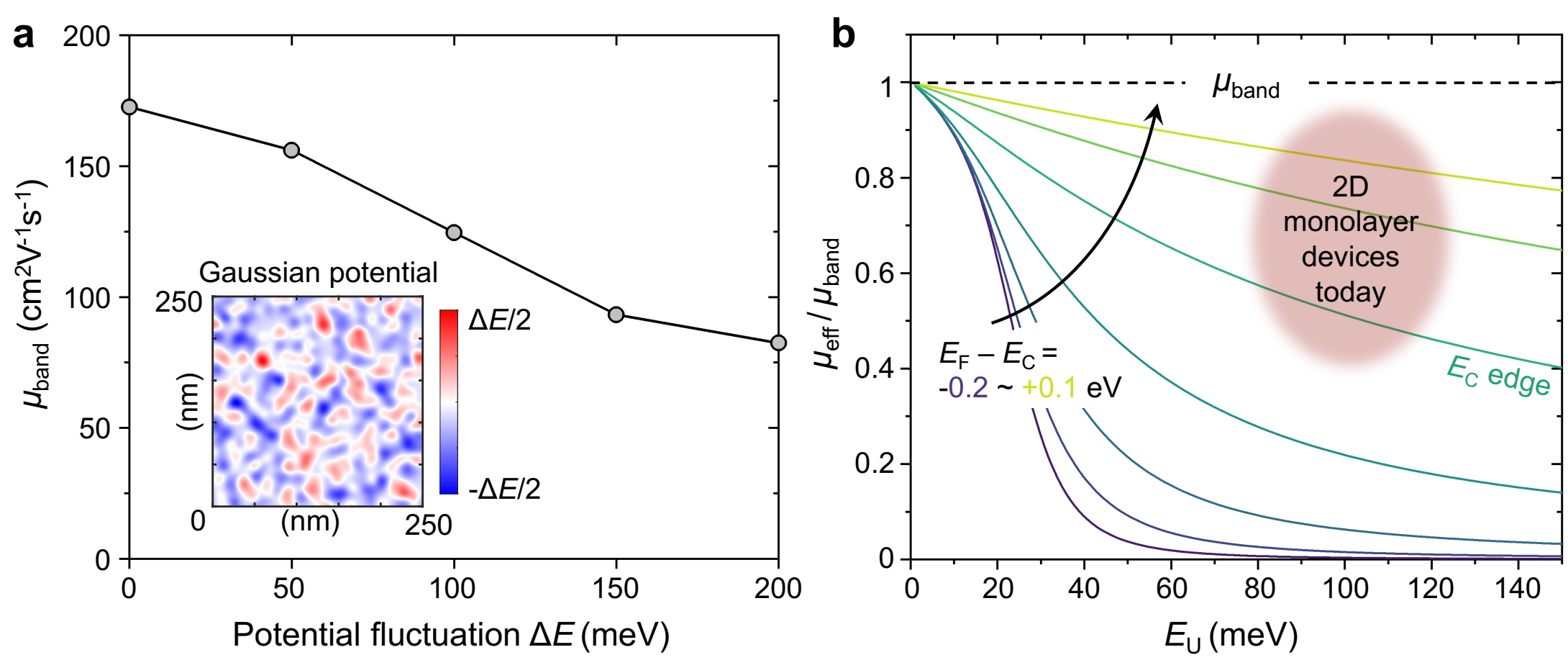

**FIG. 3. Effects of potential fluctuation and BTS on mobility. (a)** Band mobility $\mu_{band}$ as a function of potential fluctuation $\Delta E$, in monolayer $MoS_2$. This is a Monte Carlo simulation at room temperature for electron transport under a spatially-random Gaussian potential (inset shows a 250 nm × 250 nm domain). $\Delta E = 0$ corresponds to the intrinsic phonon-limited mobility reported in our previous work [28]. **(b)** Ratio of effective mobility $\mu_{eff}$ to band mobility $\mu_{band}$, as a function of $E_U$, at room temperature. Here, we assume $N_U$ changes linearly with $E_U$ from zero, and $N_U = 7.5\times10^{12}$ cm$^{-2}$ at $E_U = 100$ meV from electrical model fitting. The Fermi level $E_F$ is referenced with respect to the conduction band edge, $E_C$, and seven curves are displayed from $E_F - E_C = -0.2$ eV (purple) to +0.1 eV (light yellow) in 0.05 eV steps. The shaded area marks the approximate range of monolayer 2D devices today, gated strongly in the on-state, with $E_F \geq E_C$ corresponding to $n_{band} \geq 7\times10^{12}$ cm$^{-2}$. The $\mu_{eff}$ range is an upper limit for the mobility that may be observed experimentally, because experimental devices also have charged impurities, which are not included in this model.

Next, we investigate carrier mobility under potential fluctuations that lead to BTS in monolayer $MoS_2$ (see **Fig. 3**). We modified a previous Monte Carlo (MC) simulation approach [28] to estimate the band mobility $\mu_{band}$ for electrons [**Fig. 3(a)**], then used an MTR approach [37,43] to estimate the effective mobility $\mu_{eff}$ that is usually observed in electrical measurements of a field-effect transistor [**Fig. 3(b)**]. In the presence of a fluctuating potential, the electric field becomes nonuniform and creates additional electron scattering, lowering $\mu_{band}$. To capture this in MC simulations we include a spatially-random potential with a Gaussian distribution and peak-to-peak variation $\Delta E$, which leads to exponential BTS [13,16] (see details in Supporting Information Section V). We can estimate based on STS results [17,29] that $E_U \sim \Delta E/2$ if we take $\Delta E$ as the conduction band edge fluctuations. The simulations in **Fig. 3(a)** indicate that with 200 meV peak-to-peak potential fluctuation ($E_U \sim 100$ meV) the electron band mobility drops to nearly half of the intrinsic phonon-limited mobility [28].

**Figure 3(b)** presents the ratio between $\mu_{eff}$ and $\mu_{band}$ with varying $E_U$. Because hopping transport in BTS has negligible mobility (see Supporting Information Section II), we use an MTR-like treatment to estimate the effective mobility as $\mu_{eff} = G_{ch}/(qn_{tot}) \approx \mu_{band}n_{band}/n_{tot}$, where $G_{ch}$ is the channel conductance, $n_{tot}$ is the total gate-induced electron density, and other variables are defined earlier. **Figure 3(b)** shows that with higher $E_U$, more carriers are localized inside BTS and $\mu_{eff}$ gradually decreases. For typical $E_U$ estimated from existing monolayer $MoS_2$ devices [shaded region in **Fig. 3(b)**], $\mu_{eff}$ drops to nearly half of $\mu_{band}$ when $E_F$ is at the conduction band edge. This calculation suggests that conventional estimates of effective mobility from electrical measurements can greatly underestimate the true band mobility in the presence of BTS.

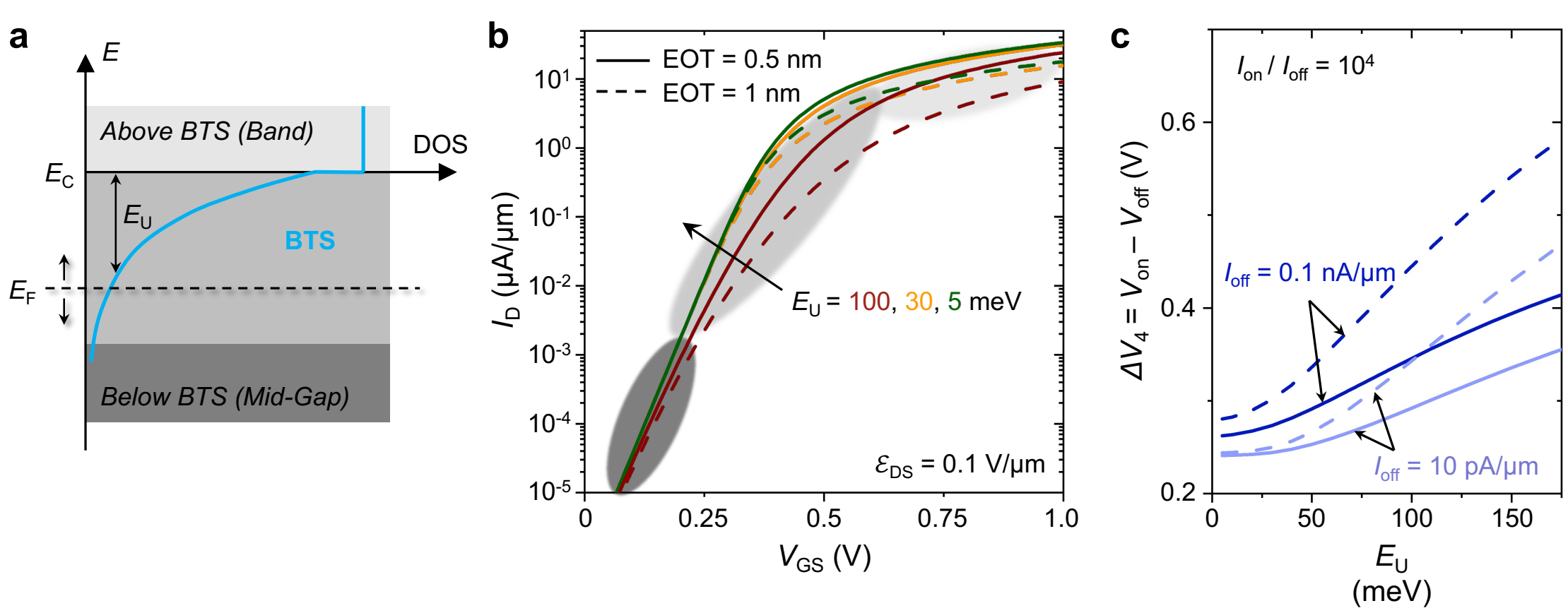

**FIG. 4. BTS effect on transistor turn-on. (a)** Energy bands and density of states (DOS, blue line) near $E_C$, including BTS. These regions are distinguished by the level of shading. Dashed line marks the Fermi level, $E_F$. **(b)** Expected transistor turn-on with three values for $E_U$ and two for EOT (0.5 nm solid lines, 1 nm dashed lines). Shaded regions correspond to the same colors in (a), as $V_{GS}$ sweeps $E_F$ through the bands. This is a long-channel device, with zero contact resistance and low lateral drain-to-source field (0.1 V/µm), such that field-assisted transport through localized states can be neglected. The band mobility $\mu_{band} = 100$ cm$^2$V$^{-1}$s$^{-1}$ here, and $N_U$ follows the same scaling as in Fig. 3. **(c)** Voltage needed ($V_{on} - V_{off}$) to turn on the transistor over four decades of current, $I_{on}/I_{off} = 10^4$. The plot shows two scenarios, with $I_{off} = 10$ pA/µm (light blue) and 0.1 nA/µm (dark blue) [44] and $I_{on}$ accordingly higher. Solid and dashed lines correspond to EOT = 0.5 nm and 1 nm, respectively.

In light of BTS, one can divide the transistor operation into three regions [see **Fig. 4(a)**], with the Fermi level $E_F$ sweeping through them, tuned by the gate voltage $V_{GS}$. The corresponding regions as the transistor turns on are shown on the $I_D$ vs. $V_{GS}$ plot in **Fig. 4(b)**. Below BTS, the transistor is operating in the deep subthreshold region and is in the off-state. As $V_{GS}$ increases, $E_F$ enters the BTS region, the transistor is near threshold and gradually turns on. In this region, localized BTS states are being filled with electrons, which have minimal contribution to current flow. After BTS states are filled at higher $V_{GS}$, $E_F$ enters the conduction band where the majority of carriers are mobile. We note the classical notion of transistor threshold voltage [45], $V_T$, is not well-defined in this picture, because the $V_T$ obtained from linear extrapolation can be larger than that from a typical (low) off-state current. This can also be seen from the "curved knee" in Fig. 2 and was previously observed in [31,46,47].

**Figure 4(b)** compares these operation regions for different $E_U$ at low source-to-drain field, corresponding to the linear transistor operation. The largest difference in this comparison comes from the middle (BTS) region: smaller $E_U$ means steeper turn-on and lower gate voltage range needed to overcome the BTS region. The three values of $E_U$ in **Fig. 4(b)** are chosen to represent three typical scenarios: 5 meV (sharp band edge), 30 meV (e-ph limit for 2D devices), 100 meV (current 2D devices).

To quantify the effects of BTS on transistor operation, we estimate the voltage needed to turn on such a device over four decades of change in current, $\Delta V_4 = V_{on} - V_{off}$, as shown in **Fig. 4(c)**. This choice is somewhat arbitrary, and could be five or more decades of current, depending on the application. Regardless, minimizing this voltage range is critical, because the energy consumption of digital circuits is proportional to the square of the transistor voltage. For our choice of four orders of magnitude current change, the thermodynamic lower limit of $\Delta V_4$ is 240 mV (= 4 × 60 mV/decade) at room temperature. Focusing on this voltage range is more practically relevant than considering the subthreshold swing or slope (SS) [48] alone, because the SS is degraded by BTS over a wider range of current, as we see above. Such phenomena are also discussed in oxides [49,50] and cryogenic silicon devices [51,52].

In **Fig. 4(c)**, we estimate the voltage range $\Delta V_4$ as a function of $E_U$, for two choices of on-state current (as labeled) and two different EOT (0.5 nm with solid lines, and 1 nm with dashed lines). We observe the voltage range quickly degrades (increases) for any $E_U > 30$ meV when the EOT is 1 nm. However, having a smaller EOT of 0.5 nm is more forgiving due to the stronger gate-channel capacitive coupling, allowing transistor operation near its fundamental limits as long as $E_U$ can be controlled below ~50 meV. This can also be seen in **Fig. 4(b)**, where the solid green and yellow curves are nearly identical. In other words, if the magnitude of BTS in 2D monolayer semiconductors can be reduced below 50 meV, low-power transistor operation can be achieved with a sufficiently low EOT.

**Understanding and Reducing BTS**

Before concluding, we return to the origins of BTS in 2D semiconductors and some strategies to reduce the energy spread of these tail states. To understand the impact of substrate dielectric roughness on the spatial inhomogeneity of 2D semiconductors, we carried out molecular dynamics (MD) simulations [53–55] as shown in **Fig. 5**, for monolayer and bilayer $MoS_2$ supported or encapsulated by a-$SiO_2$, at room temperature [also see **Fig. 1(a)** for MD setup]. These structures are built to mimic back-gated and double-gated transistors, respectively, choosing a-$SiO_2$ here as a prototypical insulator because its MD potentials [54,55] and atomic interaction with $MoS_2$ are better known (Supporting Information Section VI).

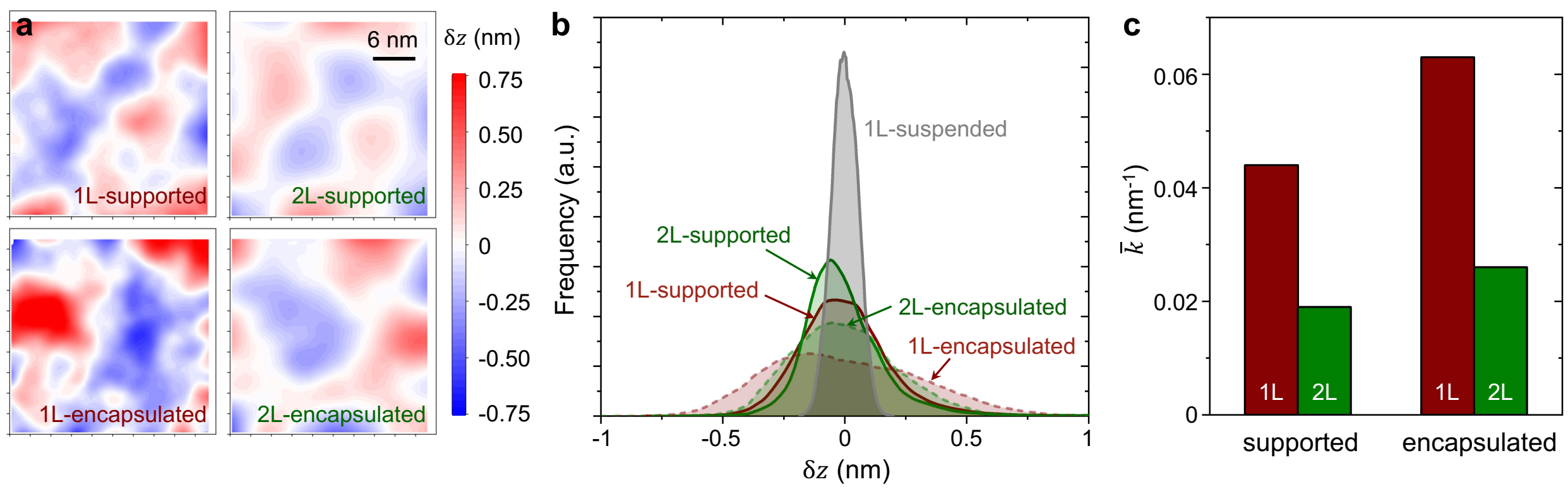

**FIG. 5. MD simulations and surface roughness. (a)** Typical surface topology from equilibrated MD snapshots. These are simulations for 1L- and 2L-$MoS_2$ films supported or encapsulated by a-$SiO_2$. The surface here is 30 nm × 30 nm. The while edge within each panel is the removed outlier data due to finite size effect of MD. **(b)** Distribution of out-of-plane atomic displacement δz for 1L- and 2L-$MoS_2$ films *supported* (solid lines) or *encapsulated* (dashed lines) by a-$SiO_2$, as well as a freely *suspended* 1L-$MoS_2$ film (grey line). The standard deviation $\sigma_z$ of these distributions corresponds to the average height fluctuations of the $MoS_2$ in the various scenarios. **(c)** The average bending curvature $\bar{k}$ of the $MoS_2$ films when supported or encapsulated by a-$SiO_2$. Monolayer in red bars while bilayer in green bars. Greater curvature (smaller radius of curvature) means greater 'wrinkling' of the atomically thin material. These $\bar{k}$ are computed by averaging $|k|$ at each local position, then averaging across multiple equilibrated MD snapshots (details in Supporting Information Section VI). All results shown are at room temperature.

We present the surface topology [see **Fig. 5(a)**] and distribution of simulated out-of-plane displacements δ*z* [see **Fig. 5(b)**] of the 2D layer when supported and encapsulated. The surface roughness is characterized by the standard deviation of this distribution, $\sigma_z$, and the average bending curvature $\bar{k}$ (see Supporting Information Section VI). The $\sigma_z$ only corresponds to height fluctuations, while $\bar{k}$ gives more comprehensive information about deformation, that can be linked to energy band fluctuations [10]. The MD results show that a freely suspended $MoS_2$ monolayer with only thermal fluctuations has $\sigma_z \approx$ 54 pm. For a monolayer supported on a-$SiO_2$, the roughness increases to $\sigma_z \approx$ 162 pm, while the roughness of the underlying substrate is $\sigma_z \approx$ 270 pm. Such numbers are consistent with literature reports [56–59] and our own atomic force microscopy measurements (AFM, details in Supporting Information Section VII).

When the 2D semiconductor is encapsulated by a-$SiO_2$ [dashed lines in **Fig. 5(b)**], we find that both monolayer and bilayer structures have higher surface roughness ($\sigma_z \approx$ 292 pm and 199 pm, respectively) than the supported cases. This may at first appear surprising, but can be understood because the top and bottom insulators have random and uncorrelated roughness profiles due to their amorphous nature. The atomically thin 2D material interacts with both top and bottom dielectric layers and can adapt to the roughness profiles of both, causing higher curvature configurations when encapsulated. We note that our MD simulations do not capture factors like the conditions of material deposition or the effect of nearby materials, like contacts, which may induce strain [35,60]. However, our results here call for careful evaluation of double-gate or gate-all-around transistors in terms of their channel surface roughness, which is extremely relevant for the development of this technology.

We also display the average bending curvature, $\bar{k}$, of supported and encapsulated $MoS_2$ in **Fig. 5(c)**. Like the surface roughness $\sigma_z$, the bending curvature is also larger for encapsulated films, although to a lesser extent in bilayer $MoS_2$. This atomic-scale bending of ultrathin 2D semiconductors causes fluctuations of their band edge energy, which can be estimated based on prior STS measurements [10] of monolayer $MoS_2$ (see Supporting Information Section I). Thus, our calculated bending curvature of supported 1L-$MoS_2$ ($\bar{k} \approx 0.044$ nm$^{-1}$) approximately yields $E_U \approx 120$ meV, which is in good agreement with our estimates from electrical measurements and modeling in **Fig. 3**. In contrast, for supported 2L-$MoS_2$ the average bending curvature is smaller ($\bar{k} \approx 0.019$ nm$^{-1}$) leading to $E_U \approx 51$ meV. Such estimates will need additional validation from future large-scale DFT and electrical measurements. Nevertheless, our results show that surface roughness is one of the major causes of BTS in monolayers, but a lesser effect in bilayers of 2D semiconductors.

The mechanical resistance of a 2D material to bending or deformation can be described by its flexural rigidity or bending stiffness $D$, which is proportional to the bending radius and inversely proportional to the bending curvature $\bar{k}$. Based on elastic plate theory [61], the flexural rigidity $D$ of an isotropic film is proportional to the cube of its thickness $h$, or $D \sim h^3$. In other words, a bilayer of $MoS_2$ may be expected to be ~8× stiffer than a monolayer of $MoS_2$. In practice, for 2D material bilayers or multi-layers, this relation is softened due to interlayer sliding, but $D$ still scales with the number of layers [62]. Ultimately, a bilayer is more rigid than a monolayer of $MoS_2$, being less prone to structural change and fluctuating potential when supported or encapsulated. This rigidity comparison explains our MD results which show smaller surface roughness for bilayer $MoS_2$ in **Fig. 5**.

We also wish to discuss potential strategies to reduce the energy spread of BTS in 2D semiconductors. We have argued that BTS can be caused by substrate surface roughness, either directly through the change of local electronic structure or indirectly through the change of local dielectric environment. Based on this reasoning, the optimal way to achieve low $E_U$ is to fabricate mechanically flat and chemically clean insulator surfaces, e.g., through interface treatment or by using crystalline insulators. However, adopting flat crystalline substrates, like $h$-BN, is generally challenging because they are less compatible with semiconductor industry requirements due to their high deposition temperature and low dielectric constant – compared with amorphous oxides such as $HfO_2$ or $ZrO_2$.

Monolayer 2D semiconductors remain electrostatically most relevant for transistors, because they represent the ultimate scaling limit of the material, and the best-case scenario for gate control. Improving the flatness of a monolayer 2D semiconductor could also be achieved by lateral tensile strain or vertical compression. Intuitively, tensile strain or vertical compression can smooth out local corrugation, thus reducing band edge fluctuations. Such strain techniques have been demonstrated in silicon [63] and recently in monolayer 2D transistors [35], but in the context of boosting band mobility of charge carriers, not in reducing BTS yet.

On the other hand, there is emerging evidence that bilayer or trilayer semiconductors may offer advantages over monolayer 2D transistors [64–67], but today it is difficult to tell if these benefits come from lower roughness and lower BTS, or from improved contact resistance, which also improves with thicker layers. In fact, bilayer 2D semiconductors may offer a sweet spot in terms of transistor performance, balancing lower BTS while retaining a thinner channel (~1.2 to ~1.4 nm) than can be practically achieved with silicon. On the other hand, 2D semiconductors thicker than approximately three layers are likely not relevant for highly scaled logic technology, because in thicker channels (2-3 nm or more), conventional semiconductors like silicon could be optimized to work sufficiently well.

## Conclusions

In summary, we have presented the origins and implications of band tail states (BTS) for transistor operation based on 2D semiconductors. The spread of BTS characterized by $E_U$ appears large today in monolayer $MoS_2$ (~100 meV) supported or encapsulated by amorphous insulators, due to potential fluctuations caused by surface roughness and inhomogeneity. This goes a long way towards explaining the reduced mobility often seen in monolayer $MoS_2$ with amorphous dielectrics compared to the same material encapsulated by flat crystalline dielectrics like *h*-BN, or from idealized theoretical calculations. The large BTS in monolayer 2D semiconductors also causes 'slow' transistor turn-on, which requires additional gate voltage, ultimately setting a limit for the energy (in)efficiency of these transistors. To realize the full potential of 2D semiconductor transistors, the BTS energy spread must be reduced below ~50 meV, either by using strain to flatten the 2D monolayer, or by employing 2D bilayers (which are stiffer), and combining these with ultrathin equivalent gate oxide thickness, down to ~0.5 nm. In this sense, we conclude that BTS represents a "speed bump" rather than the "end of road" for 2D semiconductor transistors. However, a key path to energy-efficient and low-power computing based on 2D semiconductors will require engineering the energy band edges to reduce band tail states.

## Acknowledgements

Z.H., R.K.A.B. and E.P. were supported by SRC SUPREME JUMP 2.0 Center, SystemX Alliance at Stanford University and the Department of Energy (DOE). R.K.A.B. also acknowledges support from the Stanford Graduate Fellowship (SGF) and NSERC PGS-D programs. M.A.W. acknowledges support from the National Science Foundation (NSF) Graduate Research Fellowship and the Shoucheng Zhang Fellowship, a Stanford Graduate Fellowship. T.P. would like to thank the NSF MPS Ascend Postdoctoral Fellowship. T.P. and E.P. acknowledge Intel Corporation. Atomic force microscopy was performed at nano@stanford (RRID:SCR_026695). D.D. is grateful to Zekun Chen for providing useful suggestions on the modeling setup and validation. Z.H. thanks Yujie Quan for sharing data of electron-phonon calculations on suspended monolayer $MoS_2$, B. G. Shin for sharing STS measurements of monolayer $MoS_2$ on graphite, and Wei Cai and Tony Heinz for discussions.